\documentclass[%
 aip,
 jmp,%
 amsmath,amssymb,
 reprint,%
]{revtex4-1}
\usepackage{SIunits}
\usepackage{color}
\usepackage{graphicx}
\usepackage{dcolumn}
\usepackage{bm}

\def\clr{\color{red}}
\begin{document}

\preprint{AIP/123-QED}

\title{A dual-species magneto-optical trap for lithium and strontium atoms}

\author{ Xiaobin Ma}
\author{Zhuxiong Ye}
\author{Liyang Xie}
\author{Zhen Guo}
\affiliation{
State Key Laboratory of Low-Dimensional Quantum Physics, Department of Physics, Tsinghua University, Beijing 100084, China
}%
\author{Li You}
\author{Meng Khoon Tey}
\email{mengkhoon\_tey@mail.tsinghua.edu.cn}
\affiliation{
State Key Laboratory of Low-Dimensional Quantum Physics, Department of Physics, Tsinghua University, Beijing 100084, China
}%
\affiliation{%
Collaborative Innovation Center of Quantum Matter, Beijing 100084, China
}%

\date{\today}

\def\clr{\color{red}}

\begin{abstract}

We present a machine built for experiments with ultracold mixtures of strontium and lithium atoms. The machine includes a science vacuum chamber and the relevant laser systems for cooling and trapping the atoms. With this machine, we realize a D2-line compressed magneto-optical trap (MOT) for ${^6}$Li and a narrow-linewidth 689-nm MOT for $^{84}$Sr, obtaining $\sim10^9$ ${^6}$Li atoms at 700\,$\micro$K and $\sim10^7$ ${^{84}}$Sr atoms at 1.8\,$\micro$K. Such a dual-species MOT provides an ideal starting point for realizing double degenerate mixtures of $^6$Li and Sr atoms.

\end{abstract}

\maketitle

\section{\label{sec:level1}INTRODUCTION}

Quantum gas mixtures of different chemical elements have attracted great attentions within the cold atom community. They are used to study many physics problems, such as heteronuclear Efimov resonance~\cite{2009BarontiniEfimov,2016kuhnleEfimov}, impurity in superfluids~\cite{Spiegehalder2009LiKImpurity,2010TargoImpurity,2011VernierImpurity}, and ultracold chemical reactions~\cite{2005RomanChemistry}. Of particular interest, they are employed to produce heteronuclear ground-state molecules, which can exhibit large electric dipole moments~\cite{2005SagePolarMolecules,2008DeiglmayrPolarMolecules,2008NiKRb} and thus offer interesting prospects for applications in fields ranging from precision measurement of fundamental physical constants~\cite{2012JinIntroductiontoMolecules,2008ZelevinskyMassRatio}, simulation of many-body quantum phases~\cite{2012JinIntroductiontoMolecules,2007BurcherQuantumPhase}, to quantum computation and information processing~\cite{2002DeMilleQuantumComputation}.

Most quantum-gas-mixture experiments focus on systems composed of two alkali elements~\cite{2008NiKRb,2014TakekoshiRbCs,2014MolonyRbCs,2015ParkNaKMIT,2016GuoNaRb,2017RvachovLiNa,2018BlochNaK,2019DePolarDFG}. Along with the development of cooling and trapping techniques for atoms beyond alkali elements~\cite{2003TakasuYbBEC,2005GriesmaiserCrBEC,2009KraftCaBEC,2009StellmerSr84BEC,2009EscobarSr84BEC,2011LuDyBEC,2012AikawaErBEC}, research of ultracold mixtures has now extended to richer combinations~\cite{2010TeySr87DFG,2011HaraLiYb,2011HansenLiYb,2013PasquiouRbSr,2013TakatoshiRbSr,2013BorkowskiRbYb,2015VaidyaRbYb,2016KempCsYb,2018RavensbergenKDy,2018TrantmannErDy}. Among them, mixtures of alkali and alkaline-earth(-like) atoms are of considerable interest for their possibilities of creating ground-state molecules with both an electric dipole moment and an electron spin. The open-shell molecules are proposed as promising candidates for testing fundamental symmetries~\cite{2009MeyerEDM} and for simulating lattice-spin models~\cite{2006ZollerLatticeSpin}. To date, intensive efforts have been made in investigating mixtures of Li-Yb~\cite{2011HaraLiYb,2011HansenLiYb}, Rb-Sr~\cite{2013PasquiouRbSr,2013TakatoshiRbSr}, Rb-Yb~\cite{2013BorkowskiRbYb,2015VaidyaRbYb}, and Cs-Yb~\cite{2016KempCsYb}.

In this article, we present a new apparatus aimed at producing quantum-degenerate mixtures of Li-Sr and we report a Li-Sr magneto-optical trap (MOT) realized so far. There exist a number of reasons for mixing Li and Sr. $^6$Li has been a workhorse for the studies of strongly interacting Fermi gas\cite{2010Nascimb¨¨neEOS,2012MartinEOS,2013SidorenkovSecondsound,2018Murthy2DFermiGas} and quantum simulations~\cite{2010EsslingerFermi-HubbardPhysics,2012BlochQuantumSimulations,2017GrossQuantumSimulators}. Sr possesses unique electronic transitions that allow for effective laser-cooling to quantum degeneracy~\cite{2013SimonLaserDegeneracy} and is a prime candidate for many quantum simulation proposals~\cite{2010GorshkovSU(N)magnetism,2006ZollerLatticeSpin,2008DaleyAlkalineEarthQC}. Both elements contain stable fermionic isotopes, which can be used to study fermion pairing with unbalanced mass. Ground state Li-Sr molecules exhibit an electric dipole moment of 0.33\,D (Ref.~\onlinecite{2011GopakumarAlkalineRarthLi}), nearly twice that of Li-Yb molecules. There have been a number of theoretical predictions on the properties of Li-Sr molecules~\cite{2010GueroutGroupISr,2013GuoLiSrKRb,2013GopakumarLiSr,2016PotoschnigED,2017ErikLiSr,2018ZeidSrX}, but direct observations are yet to be reported. A Li-Sr machine could fill this gap.

Wille et al. have presented the first double MOT for Li and Sr in Ref.\,\onlinecite{Wille2009}. Due to the lack of essential light sources, only a short-lived 461-nm (blue) MOT using the most abundant isotope $^{88}$Sr was demonstrated at that time. While completing this work, we have achieved a 689-nm (red) MOT for $^{84}$Sr together with a 671-nm MOT for ${^6}$Li. This article is organized as follows. In Sec.\,\ref{sec:setup}, we describe in detail the main vacuum chamber, which includes a dual-species effusive oven, a Zeeman slower, and a glass science chamber, as well as the relevant laser systems for cooling and trapping $^6$Li and $^{84}$Sr. Sec.\,\ref{sec:results} provides experimental procedure and discusses the performance of the dual-species MOT. Sec.\,\ref{sec:conclusion} concludes with a summary.

\section{\label{sec:setup}EXPERIMENTAL SETUP}

The main design considerations for this setup are to capture and cool both Li and Sr atoms from thermal gas as efficiently as possible, and to allow sufficient optical access for future experiments.


\begin{figure*}
\includegraphics[width=16cm]{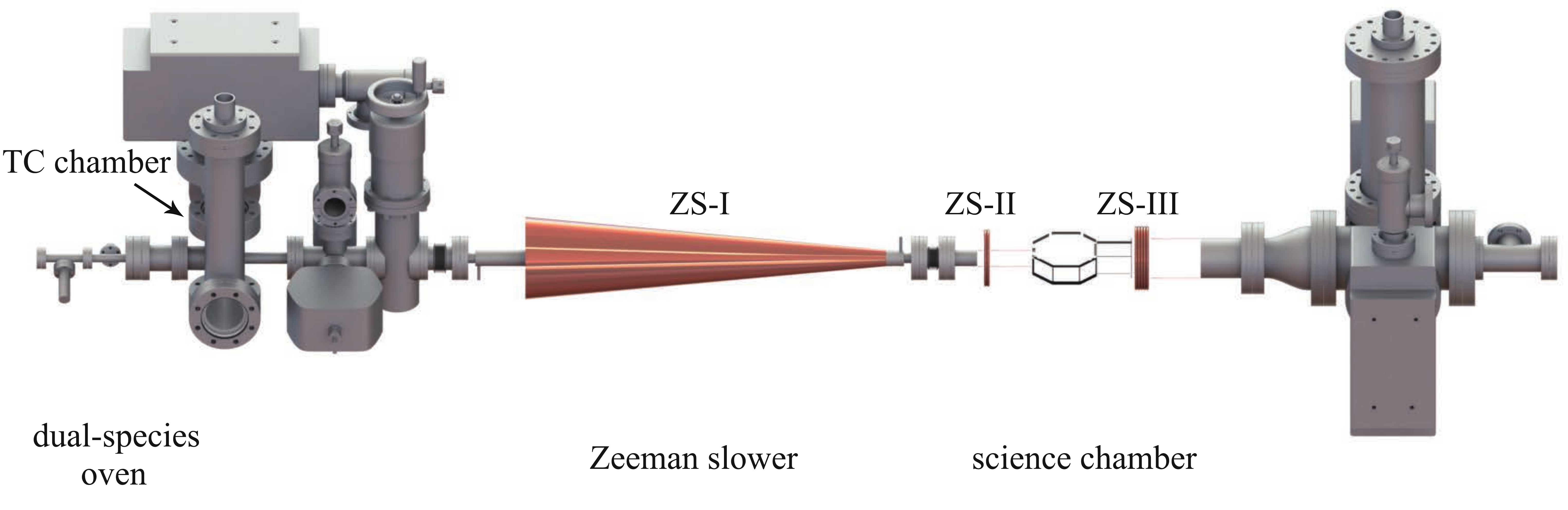}
\caption{\label{fig:LiSrMachine} 3D CAD drawings of the main vacuum system, which consists of three major parts: a dual-species oven, a common Zeeman slower and a science chamber.}
\end{figure*}

\subsection{\label{subsec:vacuum}Vacuum System}

Experiments with ultracold atoms require excellent isolation of the trapped atoms from residual gases in the vacuum. A typical vacuum pressure of order 10$^{-11}$\,Torr is required to keep the cold atoms trapped for a few tens of seconds or longer. Given the relatively low vapor pressure of Li and Sr at room temperature~\cite{1984AlcockVaporPressure}, we choose to load the dual-species MOT from decelerated atomic beams produced using a combination of high-temperature effusive ovens and a common Zeeman slower. Figure~\ref{fig:LiSrMachine} illustrates the three-dimensional computer-aided-design (3D CAD) drawing of our vacuum chamber. It includes three major parts: a dual-species oven, a Zeeman slower, and a glass science chamber. Hot Li and Sr atoms are provided by the effusive oven. They are slowed down to dozens of meters per second before being captured by a 3D MOT inside the science chamber. Our design and the projected performance for each part are presented in detail in the following.

\subsubsection{\label{subsubsec:Oven}Dual-species oven}

\begin{table}[b]
\caption{\label{tab:oven_parameters}Typical operating parameters of $^{6}$Li and $^{84}$Sr oven. }
\begin{ruledtabular}
\begin{tabular}{lcr}
 &~$^6$Li &~$^{84}$Sr\\
\hline
temperature ($^{\circ}$C) & 400 & 500\\
vapor pressure (Torr) & $1.03\times10^{-4}$ & $3.76\times10^{-3}$\\
atom density (/m$^{3}$) & $1.48\times10^{18}$ & $4.70\times10^{19}$\\
mean free path (mm) & 1802 & 30\\
most probable velocity (m/s) & 1668 & 478\\
\end{tabular}
\end{ruledtabular}
\end{table}


Different designs for a dual- or multi-species oven have been reported in the past\cite{2010OkanoLiYb,2005StanMultipleOven,Wille2009,2016KempCsYb}. Some designs mix different elements directly in a single oven. To allow for independent regulation of the Li or Sr fluxes, we adopt an design based on the Li-K-Sr oven reported in Ref.\,\onlinecite{Wille2009}.

\begin{figure}
  \centering
  \includegraphics[width=8.0cm]{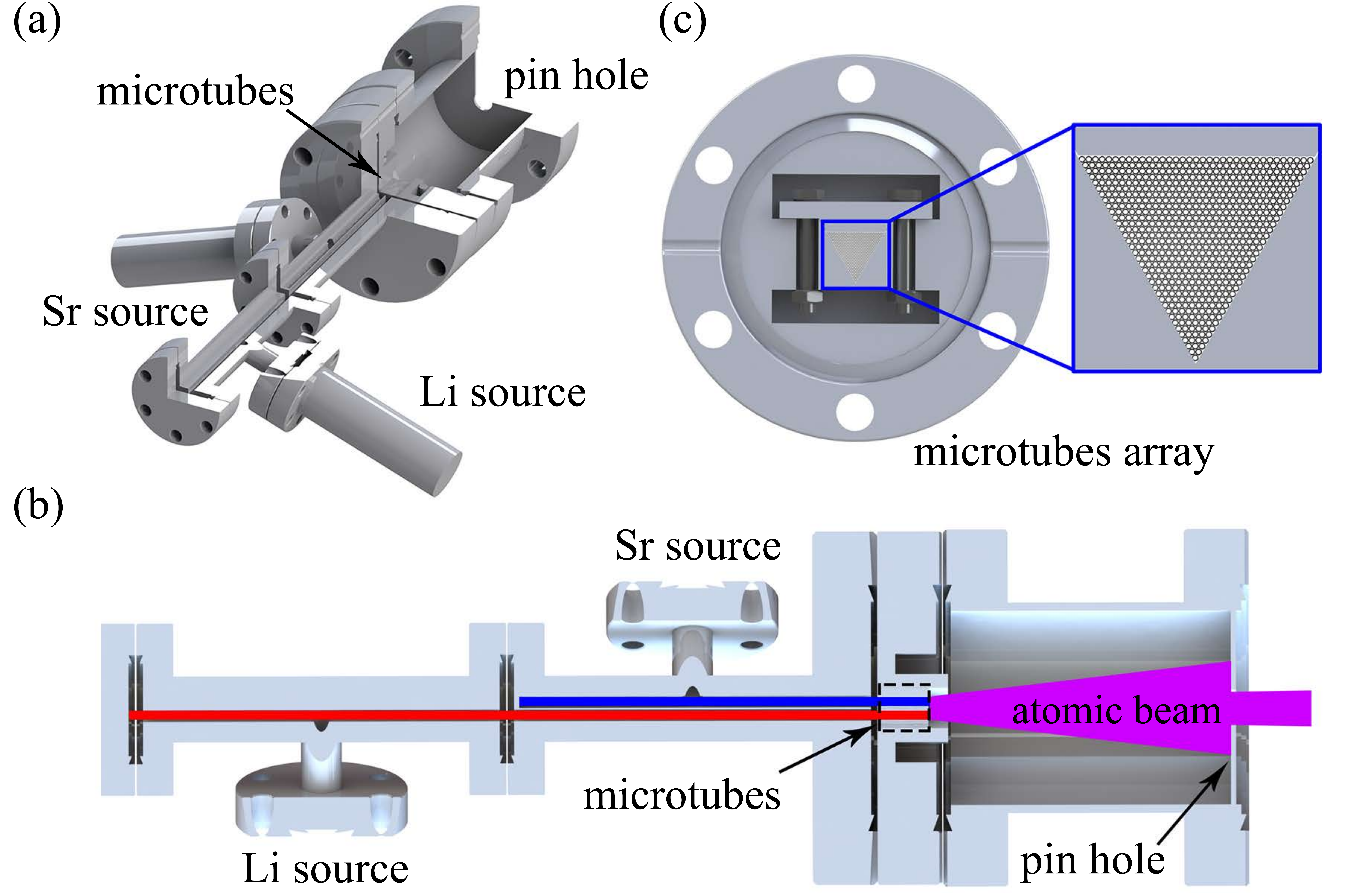}\\
  \caption{Structure of the dual-species oven. (a) Complete assembly of the oven. (b) A cutaway view illustrating the atomic beams in the oven, red for Li, blue for Sr, and purple for the mixed beam. (c) Enlarged view of the microtube array and the clamping flange. }\label{fig:oven}
\end{figure}

Our oven is composed of a Li compartment, a Sr compartment, and a nozzle, as illustrated in Fig.~\ref{fig:oven}(a). The Li and Sr compartments are almost identical in design, each consisting of an atom reservoir and a fly-through vessel connecting to the nozzle. The reservoirs are made of stainless tubes with an inner diameter of 16\,mm and a length of 55\,mm, containing 2.5\,g of 95\%-enriched~$^6$Li metal (Sigma-Aldrich, 340421) or 5\,g of Sr metal of natural abundance (Alfa Aesar, 42929). The fly-through vessel of the Sr compartment (Fig.~\ref{fig:oven}(b)) has an inner diameter of 5\,mm, and is divided into two separate channels by wire-cutting techniques, allowing the Li and Sr atoms to pass through without mixing. The Li and the Sr parts are joint via a CF16 ConFlat flange. Annealed nickel gaskets (Kurt J. Lesker, GA-0133NIA) instead of copper ones are used for sealing all the flanges exposed to Li vapor, since copper is reactive to hot Li vapor.


The divergence of the atomic beam is an important point for consideration since only atoms emitted within a limited solid angle subtended by the size of the MOT can be cooled and captured.  We adopt a popular nozzle design using an array of aligned microtubes~\cite{Wille2009,2015SenartneMicrotubeArray}, where the divergence angle of the atomic beam is determined by the aspect ratio of the microtube~\cite{1995RossAtomSources}, as long as the mean free path of atoms is larger than the tube's length (molecular flow). Our nozzle contains 600 packed stainless-steel microtubes with an outer (inner) diameter of 300\,$\micro$m (200\,$\micro$m) and a length of 10\,mm. Following the design presented by Senaratne et al.~\cite{2015SenartneMicrotubeArray}, the microtubes are packed into a triangular recess cut out of a double-sided CF35 blank flange (Fig.~\ref{fig:oven}(c)). Even though the molecular flow condition is more or less satisfied for the typical oven temperatures we use (see Table~\ref{tab:oven_parameters}), we observe substantial divergence of the atomic beam out of the oven. To prevent divergent atoms from contaminating the high-vacuum science chamber, a pin-hole with a diameter of 5\,mm is placed 6\,cm away from the nozzle (Fig.~\ref{fig:oven}\,(b)) to block these atoms.

The dual-species oven is heated using glass-fiber-tube insulated nichrome alloy wires. Its temperatures at various positions are measured with a series of K-type thermocouples and are actively controlled using commercially available PID controllers and solid-state relays. The microtube array is kept hottest to prevent clogging. The oven is thermally insulated by several layers of superwool blanket (Morgan Advanced Materials), followed by glass fiber band, and covered with aluminium foils.

The oven is connected to a transverse cooling (TC) chamber, which has four anti-reflection-coated CF63 viewports (custom-made by Larson), to provide optical access for transverse cooling beams along two orthogonal directions perpendicular to the atomic beam. The TC chamber, which is pumped by a 75\,L/s diode ion pump (Agilent, 919-1413) and a titanium sublimation pump (Agilent, 9160050), has a vacuum pressure reading of 1.5$\times$10\,$^{-9}$\,Torr at operation. Two differential pumping tubes (not shown), both with inner diameter of 5\,mm and length of 100\,mm, are used, resulting in a pressure reduction by two orders of magnitude between the science chamber and the TC chamber. The oven plus transverse-cooling section ends with a CF35 gate valve (VAT, 48132-CE01), which allows for refilling the atom sources without breaking the UHV environment in the science chamber.

\subsubsection{\label{subsubsec:Zeemanslower}Zeeman slower}

The Zeeman slower is used to generate a magnetic field profile that allows for efficient deceleration of both Li and Sr atoms down to a few tens meters per second~\cite{1982PhillipsZeemanSlower,Wille2009,2010MartiRbLiZS,2014ParisLiCs,2016BowdenLiRbZS,2016HopkinsCsYbZS}. Generally, one chooses a magnetic field profile that gives a constant deceleration $a=\eta a_{\rm max}$ over the length of the Zeeman slower, with~$\eta $ being a coefficient smaller than 1, and $a_{\rm max}=\hbar k \gamma /2m$ the maximum deceleration set by the spontaneous emission rate $\gamma$ of the transition used and $k$ the wavenumber of the slowing light, giving a magnetic field profile of the form

\begin{eqnarray}
\label{BfieldProfile}
B(z)&=&\frac{\hbar \delta}{\mu '}+\frac{\hbar kv_{0}}{\mu '}\sqrt{1-\frac{2\eta a_{\rm max}z}{v_{0}^{2}}},\nonumber\\
&\equiv&B_\mathrm{bias}+B_\mathrm{span}\sqrt{1-\frac{2\eta a_{\rm max}z}{v_{0}^{2}}}.
\end{eqnarray}
Here, $v_{0}$ is the (maximum) capture velocity of the Zeeman slower, $\mu'$ is the differential magnetic moment between the upper and lower states of the transition used, $\delta$ is the frequency detuning of the slowing light with respect to the resonance frequency of the transition, and $z$ denotes the axial position. Based on the value of $\delta$ and the sign of $\mu'$, Zeeman slower can be categorized into increasing-field slower, decreasing-field slower, and a combination of both types, called a spin-flip slower. We adopt the last type, for which the axial magnetic field decreases, crosses zero and increases again towards the direction of the MOT, 
to minimize the influence on the atoms in the MOT when the slower is switched on or off.

For practical reasons, one typically chooses a deceleration coefficient of ~$1/3<\eta <2/3$. The upper bound is chosen to avoid an over ambitious deceleration which might fail due to less than optimal conditions of the slowing light or mechanical construction, whereas the lower bound is set to prevent an overly-diverged slowed atomic beam. According to this criterion, in order to simultaneously slow down two species of atoms, the ratio between the deceleration coefficients for atom $a$ and atom $b$, $r=\eta_a/\eta_b$, should be confined to $1/2<r<2$. For a magnetic field profile given by Eq.~\ref{BfieldProfile}, it can be shown that\cite{Wille2009}

\begin{equation}
\label{ratio}
r=\frac{\eta_{a}}{\eta_{b}}=\frac{m_{a}\mu_{a} '^{2}k_{b}^{3}\gamma _{b}}{m_{b}\mu_{b}'^{2}k_{a}^{3}\gamma _{a}},
\end{equation}
which is summarized in Ref.~\onlinecite{Wille2009} for various combinations of Li, Na, K, Rb, Cs, Sr, and Yb. Among all combinations, the ratios for various isotopes of Li and Sr, which range from 1.09 to 1.33, are the closest to unity, indicating that Li and Sr are the best candidates to share a Zeeman slower. In our design, we choose $\rm \eta_{^{6}Li}=0.43$ ($\rm \eta_{^{84}Sr}=0.38$), $B\rm _{span}=1030$\,G and $B\rm _{bias}=-350$\,G. These settings result in a capture velocity of 974\,m/s and 669\,m/s for $^6$Li and $^{84}$Sr, respectively.

\begin{figure}[b]
  \centering
  \includegraphics[width=8.5cm]{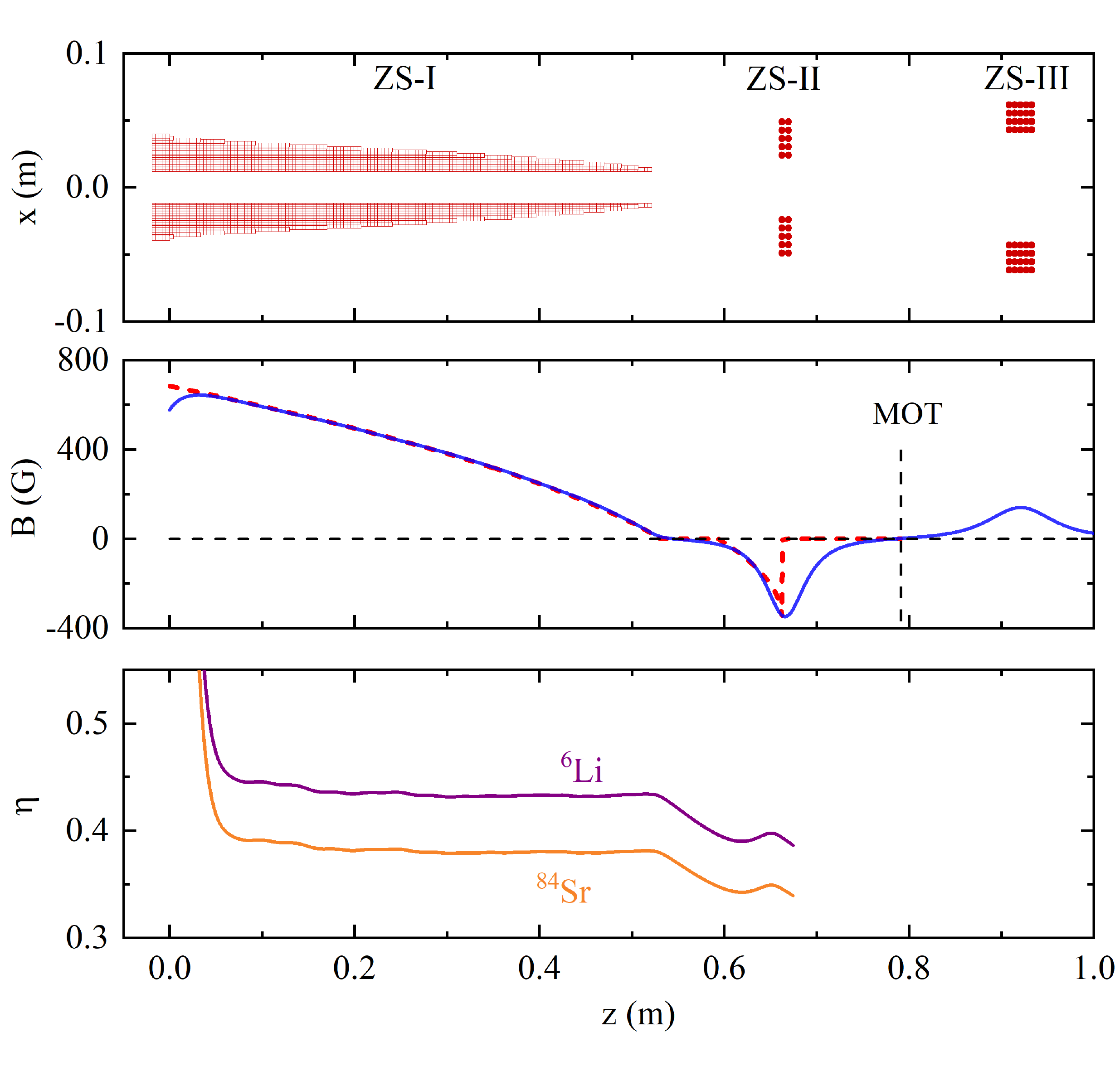}\\
  \caption{Design and specifications of our Zeeman slower. (a) Cross section view of the wire windings for the Zeeman slower. (b) The magnetic field profile along the axis of the Zeeman slower. The red dashed line shows the ideal field for uniform deceleration, while the blue line represents the simulated magnetic field for the actual coil windings. (c) The deceleration coefficients for $^{84}$Sr and $^{6}$Li at various positions of the Zeeman slower. }\label{fig:zeemanslower}
\end{figure}

The actual setup of our Zeeman slower consists of three sets of coils, ZS-I, ZS-II, ZS-III (see Fig.~\ref{fig:LiSrMachine}), with winding details shown in Fig.~\ref{fig:zeemanslower}(a). ZS-I is wound around a custom-made double-wall stainless tube, which has an outer (inner) diameter of 25.4\,mm (16\,mm). The thickness of each wall is 0.5\,mm. The space between the two walls allows cooling water to flow through. Polymide enamelled copper wires with rectangular cross section of 1.5\,mm$\times$3.5\,mm are used for winding ZS-I. ZS-II is wound around the glass to metal adaptor of the science chamber (Fig.~\ref{fig:cellwithcoil}), as close to the MOT as possible, in order to reduce the transverse expansion of the slowed atomic beam. It consists of ten turns of 4-mm-diameter hollow cooper tubes in which cooling water can flow. ZS-III consists of 20 turns of hollow copper tubes wound in the opposite side of the science chamber to compensate for the residual effects of ZS-II to the MOT. To achieve the designed field profile, ZS-I and ZS-II are operated at currents of 11.5\,A and 210\,A, respectively. The simulated on-axis magnetic field and deceleration coefficients of $^{6}$Li and $^{84}$Sr at various positions of the Zeeman slower are shown in Fig.~\ref{fig:zeemanslower}(b) and Fig.~\ref{fig:zeemanslower}(c), respectively.

Optical wise, we adopt a reflective Zeeman slower\cite{2010MartiRbLiZS} which features a flat cooper mirror coated by gold (Kugler) angled at 45$^\circ$ with respect to the atomic beam. The slightly focused slowing light beams for Li and Sr enter the vacuum system perpendicular to the atomic beam and travel against it upon reflection by the cooper mirror. This reflective scheme avoids coating of the atoms on the entry window for light, a problem which requires heating of the window (which inadvertently causes outgassing and worse vacuum) to mitigate. We have since operated the system for more than a year without observing obvious problems resulting from coating of the reflective mirror by Li and Sr.

\subsubsection{\label{subsubsec:cell}Science chamber}

\begin{figure}[b]
  \centering
  \includegraphics[width=8.5cm]{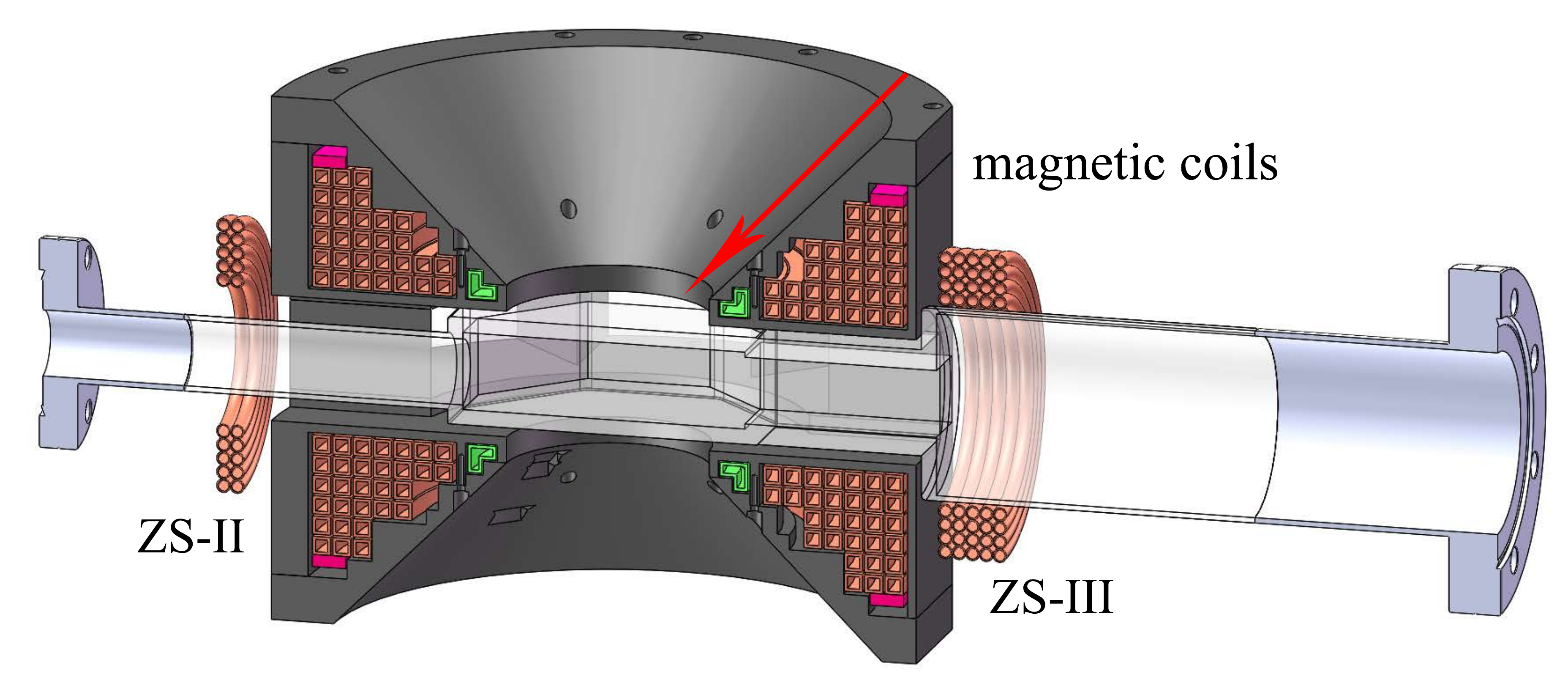}\\
  \caption{A cutaway view of the octagonal glass cell and the coil system around it. The MOT/Feshbach coils, fast switching coils, and vertical earth field compensating coils are indicated in brown, green and pink, respectively. The red arrow illustrates a light beam entering the glass cell at Brewster angle.}\label{fig:cellwithcoil}
\end{figure}

Our science chamber is a custom-made octagonal glass cell by JapanCell (Fig.~\ref{fig:cellwithcoil}). The main part of the cell is constructed from 5\,mm-thick synthetic-quartz plates joined via epoxy adhesion bonding. Dimensions of the cell are chosen to allow for a MOT with trapping volume of about 1 cubic inch. A CF35 flange and a CF63 flange made by nonmagnetic 316 stainless steel are attached to the cell, and connected, respectively, to the Zeeman slower and vacuum pumps. Compared to  metal chamber, a glass cell is free from magnetization and eddy current, thus allowing more precise and faster control of the magnetic field. It also allows for better optical access typically.

Unfortunately, the glass cell we use has a design flaw. The octagonal chamber is assembled by placing two octagonal plates at the top and bottom of 8 vertical rectangular plates. After evacuation, the atmosphere exerts approximately 1000\,N of force on both the top and bottom octagonal plates which in turn squeeze the vertical plates, resulting in serious stress-induced birefringence on the rectangular plates. On these plates, we measure a retardation of 10.0\,$^{\circ}$ (18.5\,$^{\circ}$) between the horizontal and vertical polarizations for 671-nm (461-nm) light. To ensure that the atoms see circularly polarized light for the MOT, we compensate for the polarizations of the incident Li and Sr MOT beams before they enter an achromatic quarter waveplate (Casix) placed before the glass cell.

The science chamber is evacuated by a pair of 75\,L/s ion pumps plus a titanium sublimation pump, via a CF63 flange. Pressure of the UHV region is measured using an ionization gauge (Agilent, UHV24P) to be 2.4$\times$10$^{-11}$\,Torr, which is consistent with the value estimated by measured lifetime of magnetically trapped atoms.

\subsubsection{\label{subsubsec:Coils}Coil system}

A versatile magnetic coil system, which includes MOT/Feshbach coils (switched by H-bridge), fast switching coils and weak vertical field compensating coils, is built around the science chamber, as illustrated in Fig.~\ref{fig:cellwithcoil}. The coils are housed and glued in a custom-made casing made of glass-fiber reinforced Nylon (PA66-GF30), which features a distinctive cone-shaped cut-out to facilitate optical access to the glass cell at Brewster angle (red arrow in Fig.~\ref{fig:cellwithcoil}). The MOT/Feshbach coils are able to produce magnetic field gradient up to 160\,G/cm or bias field up to 1000\,G.

\subsection{\label{subsec:laser}Laser Systems}

\begin{figure*}
  \centering
  \includegraphics[width=18cm]{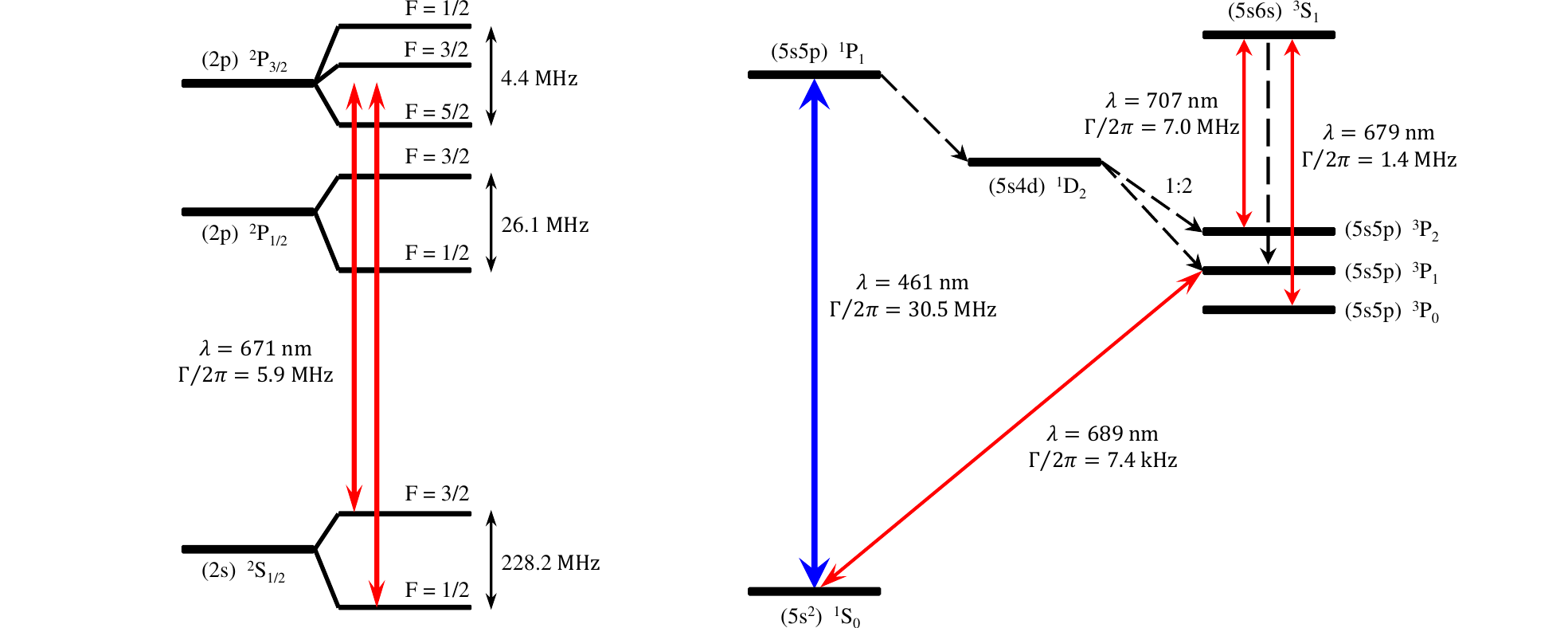}\\
  \caption{Energy level diagrams of $^{6}$Li and Sr (not to scale). The red and blue arrows represent transitions we use in the experiment for cooling $^{6}$Li and Sr. The hyperfine splitting for the $(2p)\,^{2}$P$_{3/2}$ of Li is smaller than the D2 transition linewidth (5.9\,MHz).}\label{fig:LiSrLevels}
\end{figure*}

Several laser systems are required to cool the Li and Sr atoms. Figure~\ref{fig:LiSrLevels} shows the relevant energy levels and transitions we use for cooling $^{6}$Li and Sr. For $^{6}$Li, we use the D2 transition at a wavelength of 671\,nm for both cooling and repumping. For Sr, the 461-nm $^{1}$S$_{0}\rightarrow ^{1}$P$_{1}$ singlet transition has a natural linewidth of 30.5\,MHz while the 689-nm $^{1}$S$_{0}\rightarrow ^{3}$P$_{1}$ triplet transition is four orders of magnitude narrower, this feature allows for a two-stage magneto-optical trapping of Sr to reach a temperature lower than 1\,$\micro$K~\cite{1999KatoriSrRedMOT}. Two repumping lasers at 679\,nm and 707\,nm are also used for Sr. Detailed descriptions of the light sources for $^{6}$Li and Sr are presented in the following subsections.

\subsubsection{\label{subsubsec:LiLaser}Lithium}

\begin{figure}
  \centering
  \includegraphics[width=8.5cm]{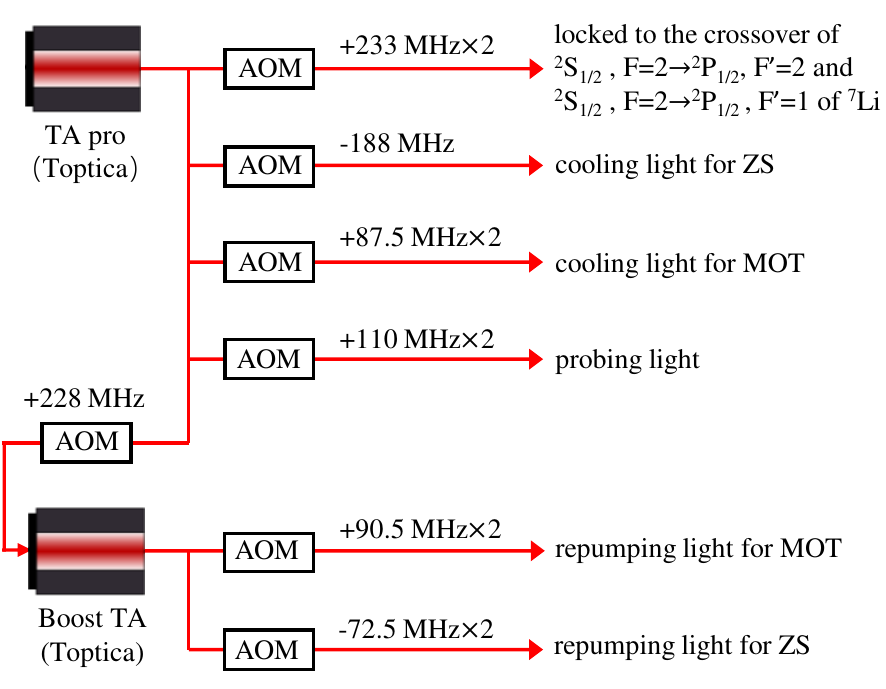}\\
  \caption{Schematics of laser systems for preparing various frequencies used for cooling and trapping $^6$Li.}\label{fig:LiFrequencies}
\end{figure}

All light needed for cooling, trapping, and probing $^6$Li atoms are generated as shown in the schematic setup of Fig.~\ref{fig:LiFrequencies}. A homemade spectroscopy heat pipe containing lithium with natural abundance provides the absolute frequency reference for our 671-nm laser systems.

\subsubsection{\label{subsubsec:SrLaser}Strontium}

The 461-nm blue light, which corresponds to the $^{1}$S$_{0}\rightarrow^{1}$P$_{1}$ transition (Fig.~\ref{fig:LiSrLevels}), is provided by a frequency-doubled diode laser system (Toptica SHG-Pro). Its frequency is stabilized using saturated absorption spectroscopy of $^{88}$Sr with the Pound-Drever-Hall (PDH) locking technique. The single laser is used for transverse cooling, Zeeman slowing, blue MOT and probing of Sr atoms after appropriate frequency shifts using acousto-optic modulators (AOMs).

As can be seen in Fig.~\ref{fig:LiSrLevels}, the $^{1}$S$_{0}\rightarrow^{1}$P$_{1}$ transition is not completely closed. During blue MOT stage, roughly one out of $600000$  (Ref.\onlinecite{Bayerle2017}) atoms in the $^{1}$P$_{1}$ state decays to the dark metastable $^{3}$P$_{2}$ state via the $^{1}$D$_{2}$ state. To retrieve the atoms trapped in the $^{3}$P$_{2}$ state, a variety of repumping schemes have been established\cite{Kurt1999,2014SimonSrRepump,2009MickelsonSrRepump}. We choose the approach that pumps the atoms to the $^{3}$S$_{1}$ state because the required lasers, 707\,nm (679\,nm) for $^{3}$P$_{2}\rightarrow^{3}$S$_{1}$ ($^{3}$P$_{0}\rightarrow^{3}$S$_{1}$) transition, are readily available (Opnext, HL7001MG and HL6748MG). The frequencies of the two repumping lasers are monitored by a wavelength meter (HighFiness, WS-6) without feedback control.

Preparation of the 689-nm light is more challenging, because at 7.4\,kHz, the $^{1}$S$_{0}\rightarrow^{3}$P$_{1}$ transition linewidth is much smaller than the linewidth ($\sim100$\,kHz) of our laser (Toptica, TA-Pro). To compress the linewidth of the laser, we lock the (AOM-shifted) frequency of the 689-nm laser to an optical cavity with a finesse of 220000 and a spacer made of ultra low expansion glass (ULE) using PDH technique. For long-term stability (since the length of the reference cavity drifts slowly over time), the actual frequency of the laser is locked to the saturated absorption spectrum of $^{88}$Sr by tuning the frequency of the AOM which connects the laser to the aforementioned linewidth-reduction cavity locking system. Using the beat signal between two 689-nm lasers locked to the same reference cavity, we estimate the linewidth of our 689-nm light to be less than 1\,kHz.

\subsection{\label{subsec:MOTsetup}MOT setup}

\begin{figure}
  \centering
  \includegraphics[width=8.5cm]{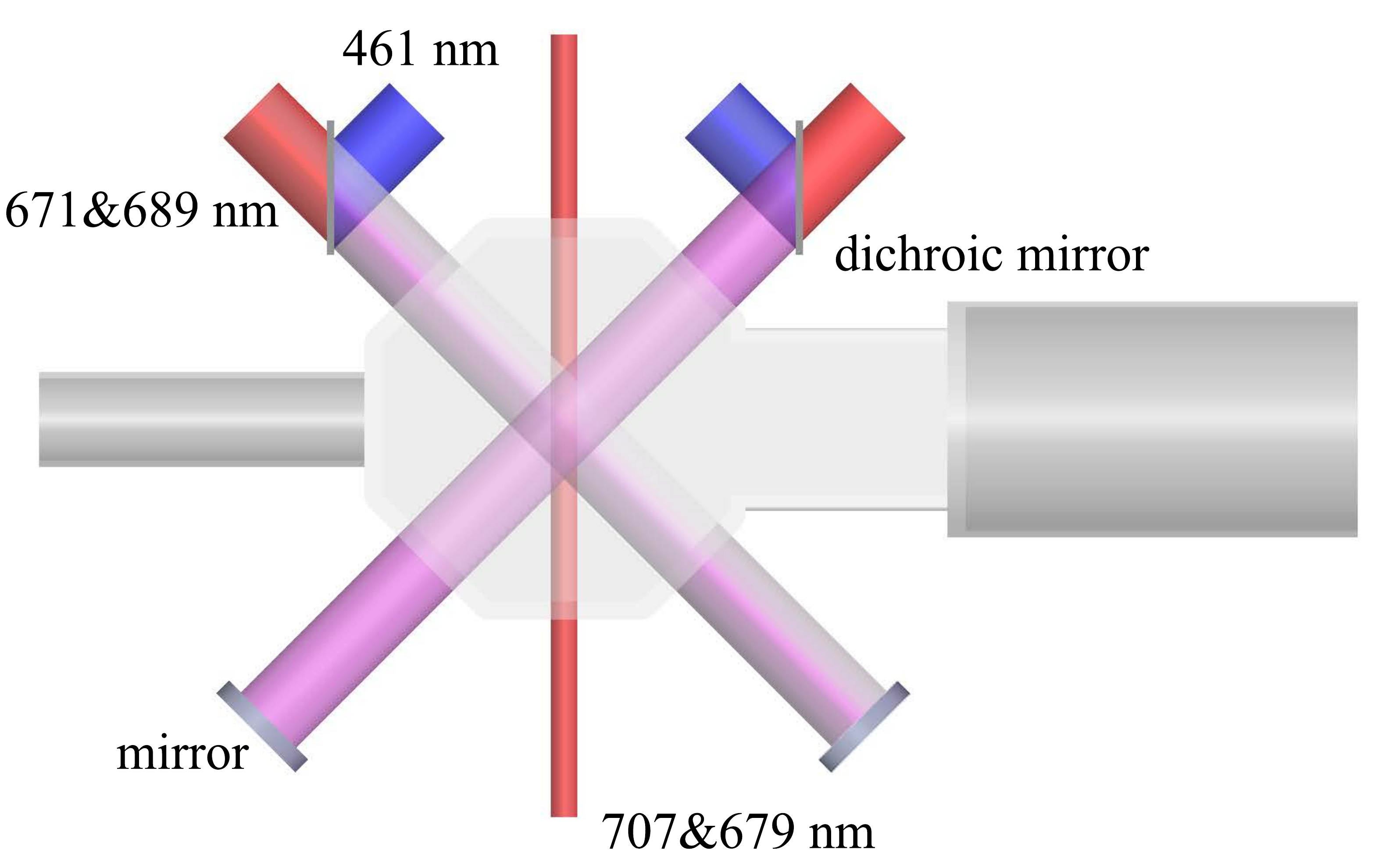}\\
  \caption{Top view of the retro-reflected MOT setup. Most optical elements are not shown for clarity. The vertical MOT beams are also not shown.}\label{fig:motsetup}
\end{figure}

As shown in Fig.~\ref{fig:motsetup}, a three-beam retro-reflected configuration is used for the MOT setup due to limited power of our light sources. To combine four different frequencies required for the dual-species MOT, we first mix the red lasers (including the 671-nm cooling and repumping lights for Li MOT, and the 689-nm light for Sr red MOT) via a 4-in-4-out polarization-maintaining (PM) fiber splitter (custom-made by Evanescent)\footnote{The spare input port of the 4-in-4-out fiber splitter is reserved for the repumping light for $^{87}$Sr red MOT or for the gray molasses lights for Li later.}. This fiber splitter splits each input equally to 4 output ports. We use only three of the outputs for the MOT setup. An independent 1-in-3-out PM fiber splitter is used to prepare three 461-nm beams for the Sr blue MOT. Finally, the three red beams and blue beams are combined using long-pass dichroic mirrors (Thorlabs, DMLP550R), as shown in Fig.~\ref{fig:motsetup}. The three resulting mixed beams overlap orthogonally at the center of the quadrupole field generated by the MOT coils. Each of them has a 1/e$^{2}$ waist of about 6\,mm and is circularly polarized after passing through an achromatic quarter-wave plate (Casix). The combination of an achromatic quarter-wave plate and a broad-band mirror is employed to retro-reflect the MOT beams. The 679-nm and 707-nm repumping light for Sr passes through the center of the MOT at 45$^\circ$ with respect to the horizontal MOT beams (Fig.~\ref{fig:motsetup}).

\section{\label{sec:results}Results}

In this section, we report the performance of our MOTs. The number and temperature of the trapped atoms are measured via the standard absorption imaging technique using an Andor iKon M934 camera. Sr atoms are imaged using the 461-nm transition, while Li atoms are imaged using the D2 line at 671-nm.

\subsection{\label{subsec:LiMOT}Lithium MOT}
\begin{figure}
  \centering
  \includegraphics[width=8.5cm]{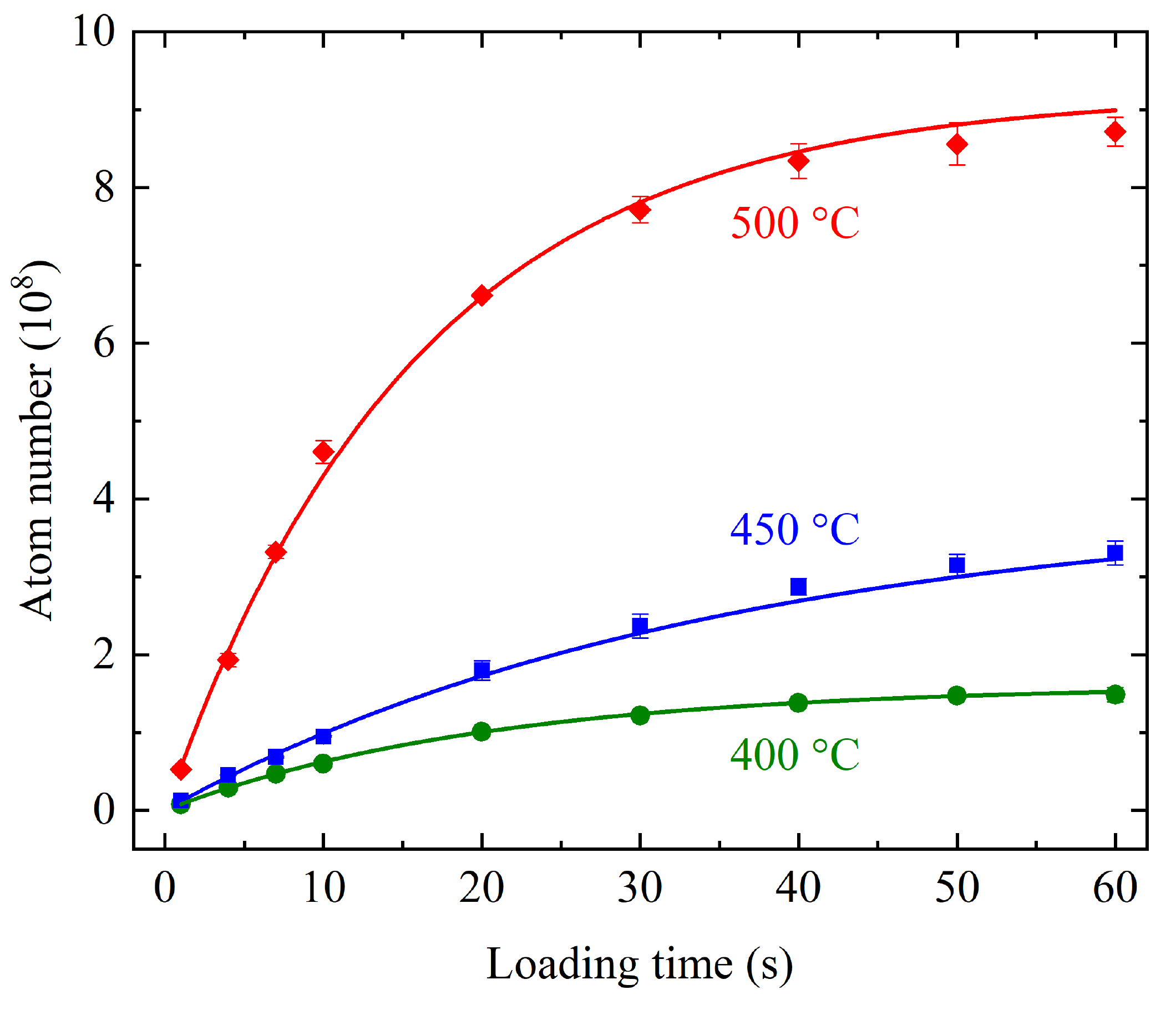}\\
  \caption{Number of $^6$Li atoms in the MOT vs loading time for different oven temperatures.}\label{fig:Li6Loading}
\end{figure}

In our experiment, the magneto-optical trapping of $^6$Li undergoes two stages: a regular MOT stage for atom loading and a compressed-MOT stage for increasing the density and reducing the temperature of the atoms. In the former stage, the cooling (repumping) light is 45\,MHz (39\,MHz) red detuned from the $^2S_{1/2},F$=$3/2$ to $^2P_{3/2}$ ($^2S_{1/2},F$=$1/2$ to $^2P_{3/2}$) transition (Fig.~\ref{fig:LiSrLevels}) with a power of 9.5\,mW (3.0\,mW) in each beam. The magnetic field gradient in the vertical direction is optimized to be 20\,G/cm. Once the loading is finished, the MOT is compressed by linearly ramping up the magnetic field gradient to 38\,G/cm in 3\,ms. Meanwhile, the power and the frequency detuning of the cooling (repumping) light are reduced to 0.26\,mW(45\,$\micro$W) and -6\,MHz (-6\,MHz), respectively. Figure~\ref{fig:Li6Loading} shows the loading curves of $^{6}$Li MOT for various temperatures of the Li oven. For an oven temperature of 500\,$^{\circ}$C, we are able to obtain 4.5$\times$10$^{8}$ $^6$Li atoms after 10\,s loading, with a typical temperature of 700\,$\micro$K.

\subsection{\label{subsec:SrMOT}Strontium MOT}

\begin{figure}[h]
  \centering
  \includegraphics[width=8.6cm]{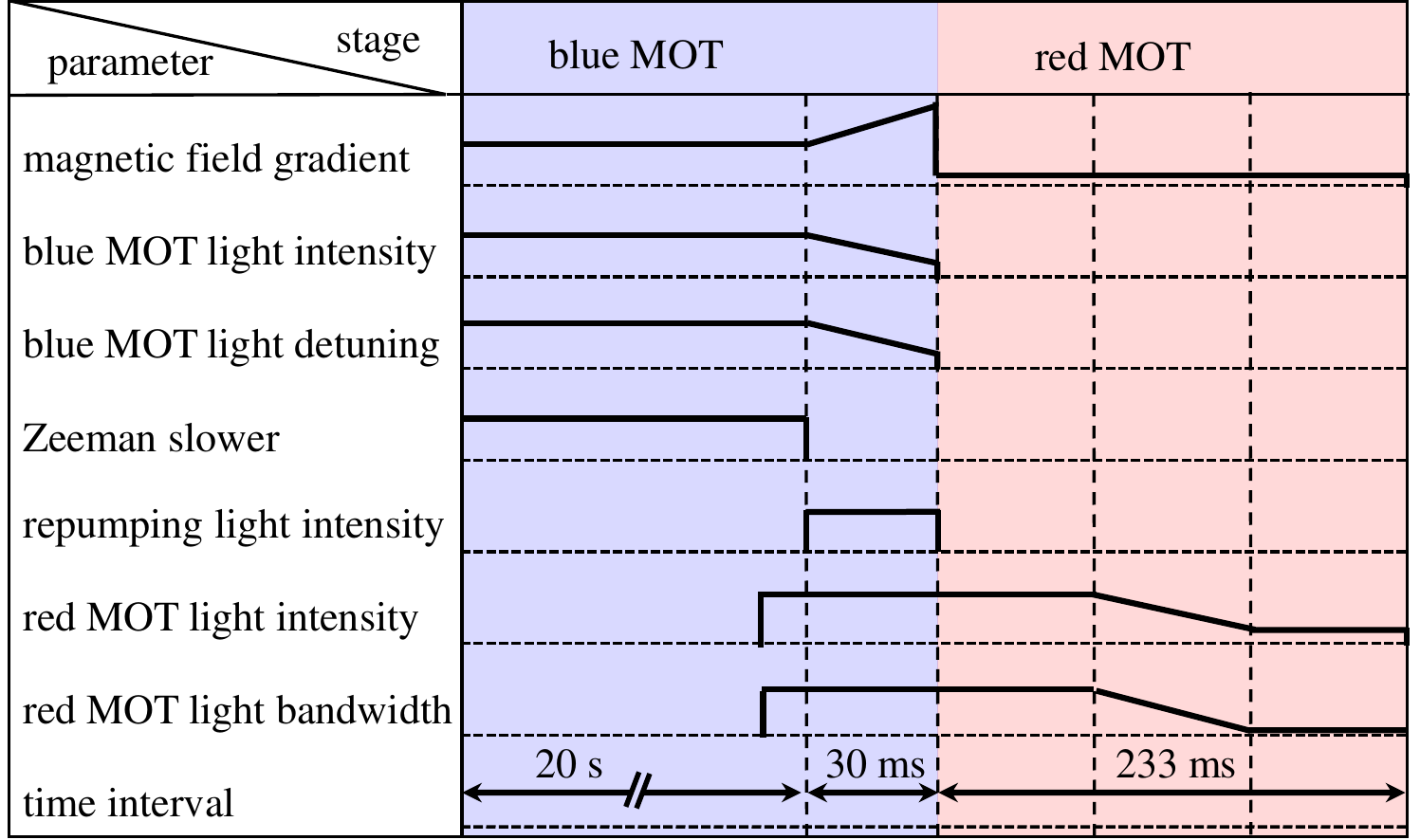}\\
  \caption{Schematics of experimental sequence for the blue and red MOTs of $^{84}$Sr. }\label{fig:srsequency}
\end{figure}

Figure \ref{fig:srsequency} shows the experimental sequence for preparing the blue and red MOTs of $^{84}$Sr. Atoms precooled by Zeeman slower are first captured by the blue MOT and then transferred to the red MOT after a repumping stage in between. During loading, the blue MOT laser beams have a detuning of -40\,MHz, and each beam contains a power of 4.0\,mW. The magnetic field gradient along vertical direction is optimized to be 51\,G/cm. By accumulating Sr atoms in the dark metastable $^3P_2$ state (Fig.~\ref{fig:LiSrLevels})~(Ref.~\onlinecite{2003NagelSrMagneticTrap}), more than 2.0$\times$10$^{7}$ of the lowest abundant $^{84}$Sr can be magnetically trapped with a loading time of 20\,s. To reduce the temperature of the blue MOT before further cooling using the red MOT, we slightly ramp up the magnetic field gradient, and reduce the detuning and power of 461-nm light to -32\,MHz and  1.6\,mW in 30\,ms, turning on the 679-nm and 707-nm repumping light at the same time (see Fig.~\ref{fig:srsequency}). The temperature of the $^{84}$Sr atoms decreases from 3\,mK to 1\,mK during this stage.

\begin{figure}
  \centering
  \includegraphics[width=8cm]{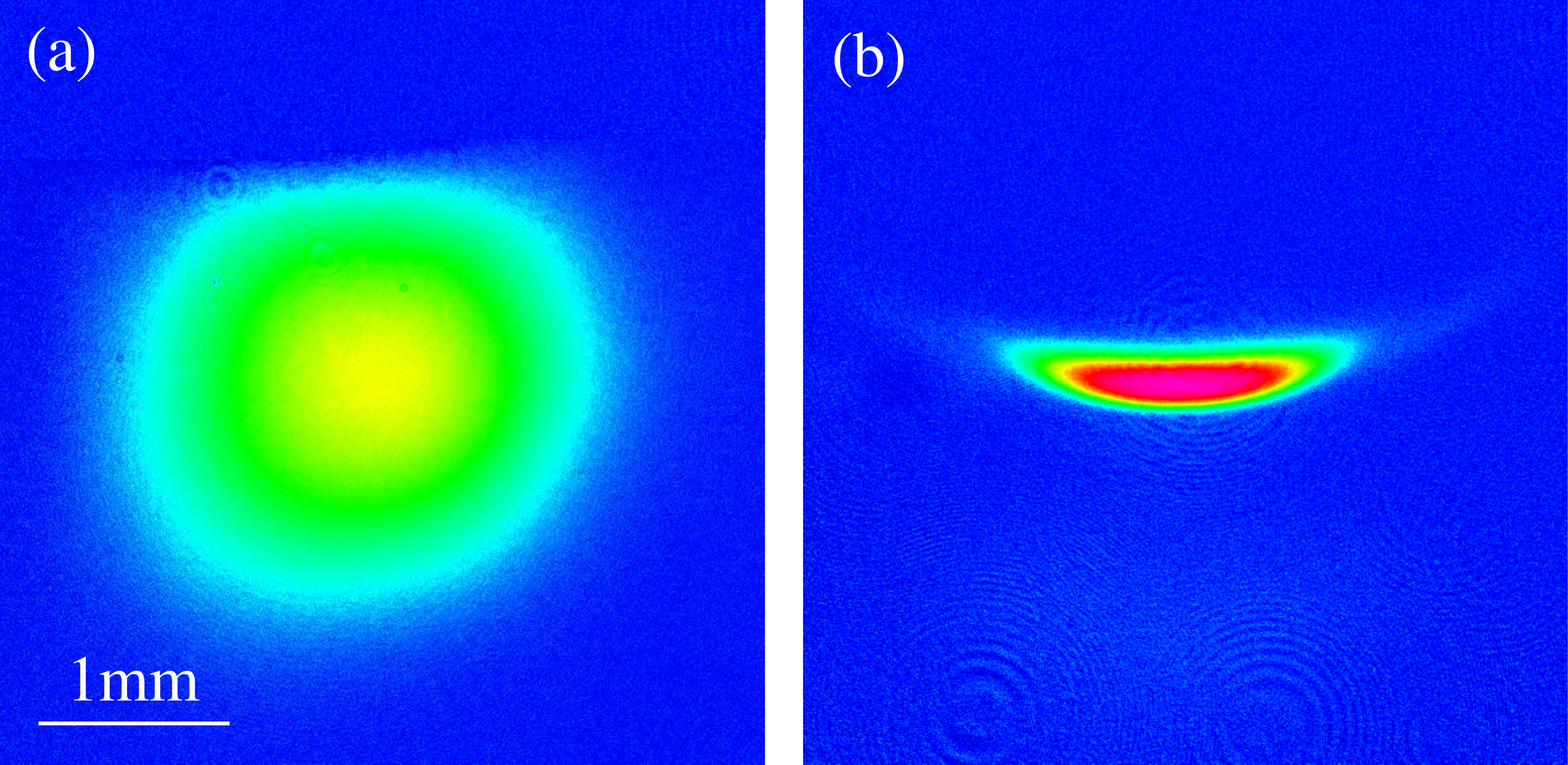}\\
  \caption{Absorption image of (a) blue MOT and (b) red MOT of $^{84}$Sr viewed along the horizontal direction. The flight time of blue (red) MOT is 0.1\,ms (0.5\,ms). The pancake shape of the red MOT arises from the balancing act between the gravity and the weak light scattering force.}\label{fig:SrMOT}
\end{figure}

Transfer of $^{84}$Sr atoms from the blue MOT to the red MOT starts by extinguishing the 461-nm light and setting the magnetic field gradient to 4.3\,G/cm. The 689-nm light for the red MOT, with a power of 9.5\,mW in each beam, is turned on 100\,ms before switching off the 461-nm light (Fig.~\ref{fig:srsequency}). To achieve a good transfer efficiency from the blue MOT to the red MOT, a broad-band red MOT is adopted by modulating the frequency of the 689-nm light. In the first 75\,ms of the red MOT plus the 100\,ms in the blue MOT stage, this modulation generates a frequency comb extending from -6.9\,MHz to -100\,kHz (with respect to the resonant frequency of the $^1S_0-^3P_1$ transition) with a comb spacing of 50\,kHz. After that, the modulation span (modulation frequency) is ramped linearly to -3.9\,MHz -- -100\,kHz (10\,kHz) in 88\,ms. Meanwhile, the power of each red MOT beam is ramped to 1.4\,mW. Finally, the detuning (power) of the red MOT beam is set to -450\,kHz (56\,$\micro$W) and held for 60\,ms. At the end of the red MOT cooling stage, over 70\% atoms from the blue MOT remain, with the atom temperature decreasing from 1\,mK to 1.8\,$\micro$K, see Fig.~\ref{fig:SrMOT}.

\subsection{\label{subsec:LiSrMOT}Simultaneous dual-species MOT}

The main challenge for realizing a simultaneous $^{6}$Li and $^{84}$Sr MOT comes from the different optimal magnetic field gradients for the two species (despite the fact that they are suitable for sharing a single Zeeman slower). Using a compromised magnetic field gradient of 35\,G/cm, the trapped atom numbers of $^{6}$Li and $^{84}$Sr drop by 51\% and 62\%, respectively. The mismatch between the optimal magnetic field gradient also makes it difficult to produce the red MOT of $^{84}$Sr and $^{6}$Li MOT at the same time. These observations suggest that it may be more convenient to load and transfer Li and Sr atoms into an optical dipole trap in a consecutive manner rather than using a simultaneous MOT.

\section{\label{sec:conclusion}CONCLUSION}

We describe a new machine for magneto-optical trapping $^{6}$Li and $^{84}$Sr atoms. Using this machine, we have obtained $\sim10^9$ ~$^6$Li atoms at 700\,$\micro$K in a D2-transition compressed MOT, and $\sim10^7$ $^{84}$Sr atoms at 1.8\,$\micro$K in a narrow-linewidth 689-nm MOT. This provides an ideal starting point for possible realization of double degenerate mixtures of $^6$Li and Sr atoms, and for experiments with ultracold mixtures of the two atomic species.
\begin{acknowledgments}
This work is supported by the National Key R\&D Program of China (Grant No. 2018YFA0306503 and No. 2018YFA0306504) and by the NSFC (Grant No. 91636213, No. 91736311, No. 11574177, No. 91836302, and No. 11654001).
\end{acknowledgments}

\nocite{*}

%

\end{document}